\documentclass[a4paper,11pt]{article}
\pdfoutput=1 

\usepackage{jinstpub} 
\usepackage{subfig}
\usepackage{upgreek}
\usepackage{lineno}
\usepackage{indentfirst}

\title{\boldmath Simulation of charge transport at the surface of planar silicon sensors}

\author[b]{I. Bloch,}
\author[b]{B. Bruers,}
\author[a]{H. Lacker,}
\author[a,1]{P. Li,\note{Corresponding author.}}
\author[b]{I.-S. Ninca,}
\author[a]{C. Scharf}

\affiliation[a]{Humboldt University of Berlin,\\Unter den Linden 6, 10117 Berlin, Germany}
\affiliation[b]{Deutsches Elektronen-Synchrotron (DESY) ,\\Platanenallee 6, 15738 Zeuthen, Germany}

\emailAdd{lipeilin@hu-berlin.de}

\abstract{ Radiation-hard silicon sensors used in high-energy physics require a high electric field and are susceptible to surface breakdown. This study aims to improve the understanding of the underlying mechanisms by developing new methods to probe the electric field at surface near the sensor’s edge. For planar sensors, avalanche breakdown primarily occurs at the Si-SiO$_2$-interface, where localized electric field peaks can form between the guard ring and the edge. Accurate simulations are challenging and it is essential to validate simulation parameters by comparing the simulation results to measurements. In this work, the electrical behavior of the edge region of planar silicon diodes was simulated using Synopsis TCAD. Transient Current Technique (TCT) simulations were performed in both TCAD and Allpix$^2$, and compared to measurements. Additionally, laser scans over the edge region were performed in Allpix$^2$ to evaluate the simulated surface electric field and charge collection efficiency.

}


\proceeding{20$^{\text{th}}$ Anniversary "Trento" Workshop on Advanced Silicon Radiation Detectors\\
  2.-4. Feb. 2025\\
  Trento}

\begin{document}
\maketitle
\flushbottom

\section{Introduction}
\label{sec:intro}

Silicon sensors are susceptible to avalanche breakdown if there are localized regions where the absolute electric field reaches values in the order of 200-300\,kV/cm~\cite{okuto_crowell}. This can easily happen at the edges of planar sensors, where the potential changes by several 100\,V over a distance of a few 100\,$\upmu$m~\cite{GR}. Electric field peaks develop especially at the tips of metal contacts or implants, which are located at the surface for planar sensors. Additionally, the local electric field at the surface is influenced by trapped charges near the Si-SiO$_2$ interface and, in the presence of ambient humidity, it can be strongly influenced by mobile charges on top of the SiO$_2$/passivation~\cite{GR, ninca2024tcad}.
Given the complex nature of surface conditions, accurate simulation for the surface electric field presents challenges. This study proposes an approach to characterize the surface electric field distribution by illuminating the top of the sensor similar to~\cite{GR}, but with focused laser pulses.

\section{Experimental methods}
\subsection{Device under test}
\begin{figure}[htbp]
\centering 
\includegraphics[width=0.8\textwidth, height =0.25\textwidth ]{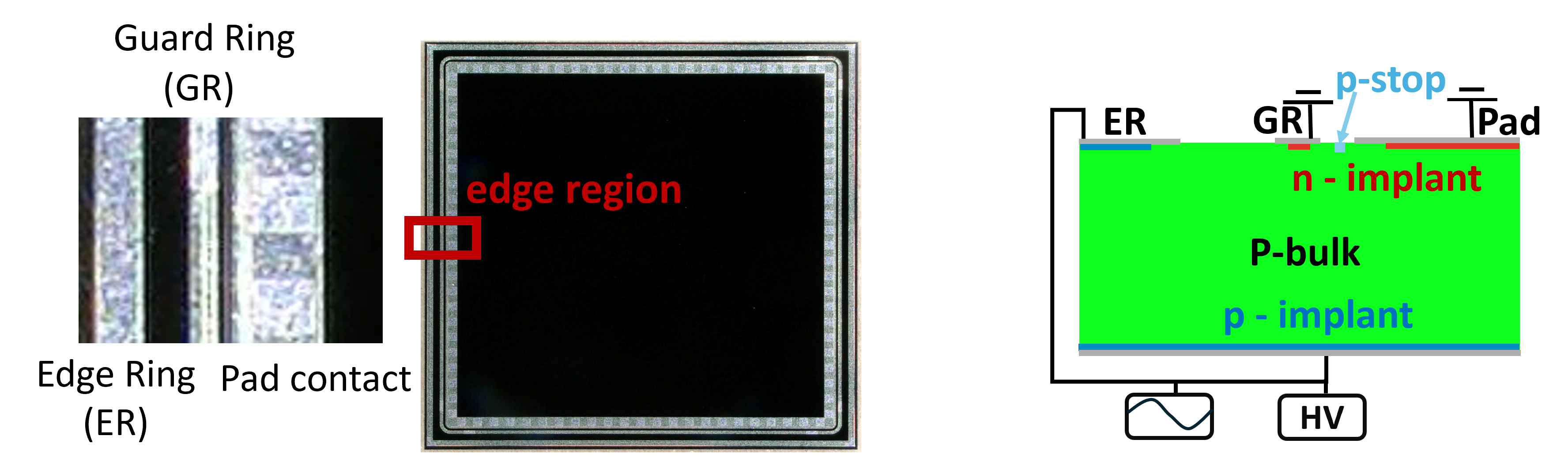}
\caption{\label{fig:setup} Picture of the $8\times8\,\text{mm}^2\,\text{n}^+/\text{p}/\text{p}^+$ mini-diode~(left) and schematic cross section of the edge region of the diode indicating the potentials at the electrodes in the TCT setup~(right).
}
\end{figure}

The edge region between the edge ring (ER) and the pad electrode of an special $8\times8\,\text{mm}^2\,\text{n}^+/\text{p}/\text{p}^+$ diode (MD8)~\cite{123} was investigated. The effective p-bulk doping concentration of around $4.2\cdot 10^{12}\,\text{cm}^{-3}$, the active thickness of around $295\,\upmu$m and the full depletion voltage of 274\,V were extracted from the capacitance-voltage measurements. The pad electrode and the guard ring (GR) are connected to ground. Between the pad and the GR there is a floating p-stop implant implemented to arrest the surface current. The ER is wire-bonded to the same node as the backside, and connected to high voltage as shown in Fig.~\ref{fig:setup}. This wire-bonding is intended to connect the ER and the backside electrode as a single unit to simplify the modeling of the weighting potential in the simulation. Transient currents are read out from the backside-ER electrode using a bias-T (not shown in Fig.~\ref{fig:setup}), which decouples the DC high voltage (HV) from the oscilloscope.

\subsection{Transient Current Technique}
The Top Transient Current Technique~(Top-TCT) was utilized to investigate surface charge transport and the surface electric field in the edge region of the MD8 diode. 

Short laser pulses ($\text{FWHM}<350$\,ps) with a wavelength of 660\,nm were focused to have the beam spot with FWHM of 10\,$\upmu$m, illuminating the front side of the diode and generating approximately 10$^4$\,electron-hole pairs. Given an absorption length of around 3.5\,$\upmu$m in silicon, most carriers were created near the surface. As the generated charge carriers drift through the diode volume, they induce transient currents in the electrodes. The initially induced current on the readout electrode, following the standard~\cite{ramo} Shockley-Ramo theorem\footnote{
Since the lateral edge of the MD8 diode is not depleted, 
a time-dependent weighting field should be used~\cite{wp}. However, only minor differences are observed comparing the TCAD simulation to the Allpix$^2$ simulation, the latter using a static weighting field (see Fig.~\ref{fig:transients2}).}, 
can be written as: $I_{ind}(t \rightarrow 0^+) \approx e_{0} \cdot N_{q} \cdot \vec{E}(x) \cdot \vec{E} _{w}(x) \cdot \sum_{e,h} \left( \mu_{e,h}(E) \right)$, where $e_{0}$ is the elementary charge, $N_{q}$ is the number of generated electron-hole pairs, $\vec{E}$ is the electric field, $x$ denotes the laser position, $\vec{E} _{w}$ is the weighting field of the readout electrode, and $\mu_{e,h}$ are the carrier mobilities. 
Accordingly, the initial "prompt" current~\cite{prompt_current} at $t \rightarrow 0^+$, right after injection of the charge carriers, contains information about the electric field at the location where the charge was generated.

\section{Simulations}

\subsection{TCAD device simulation}

Synopsys Sentaurus TCAD~\cite{tcad} was used to simulate the current, capacitance, and transients for laser illumination in the edge region of the MD8 diode. The implemented geometry of the edge region in TCAD was the same as shown in Fig.~\ref{fig:setup}.
The most relevant parameters used in the simulation are listed in Table~\ref{tab:tcad_parameters}.

The laser beam was modeled with a Gaussian temporal profile having a FWHM of 70~ps, and a Gaussian spatial intensity distribution with a FWHM of 10\,$\upmu$m. The peak intensity was lowered to 0.02\,W/cm$^2$ 
to ensure consistent results and to avoid plasma effects.

The ⟨111⟩ charge carrier mobility from Canali~\cite{canali} was used\footnote{The diode under test is from a ⟨100⟩ silicon wafer. The dicing edges are in the ⟨110⟩ direction and the ⟨111⟩ direction points 45° from the surface. Due to the
lack of a mobility model for this case, the Canali parametrization for the ⟨111⟩ direction was used, approximately aligning with the direction of the hole drift in Fig.~\ref{fig:-50line}.}, and charge multiplication was described using the Van Overstraeten–De Man model~\cite{van}.

\begin{table}[h!]
\centering
\begin{tabular}{|c|c|c|c|}

\hline
P-bulk active thickness & $295 \, \upmu\text{m}$ & SiO$_{2}$ and Si$_3$N$_4$ layer thickness & $0.6 \,\upmu\text{m}$ \\

P-bulk doping & $4.2\times 10^{12} \, \text{cm}^{-3}$ &Fixed oxide charge & $1 \times 10^{11}\,\text{cm}^{-2}$\\

Implant peak doping & $1\times 10^{19}\,\text{cm}^{-3}$&Implant and p-stop depth & $2.2 \, \upmu\text{m}$  \\
P-stop peak doping & $1 \times 10^{16}\,\text{cm}^{-3}$ & Effective lifetime $\tau_{e},\tau_{h}$ & $0.07$~s \\

\hline
\end{tabular}
\caption{Parameter values used for modeling for TCAD device simulation. }
\label{tab:tcad_parameters}
\end{table}

\begin{figure}[htbp]
\centering 
\subfloat[Absolute electric field in the edge region]{\label{fig:efield_full}
\centering
\hspace{-0.5cm}
\includegraphics[height=.155\textheight, width=.31\textheight]{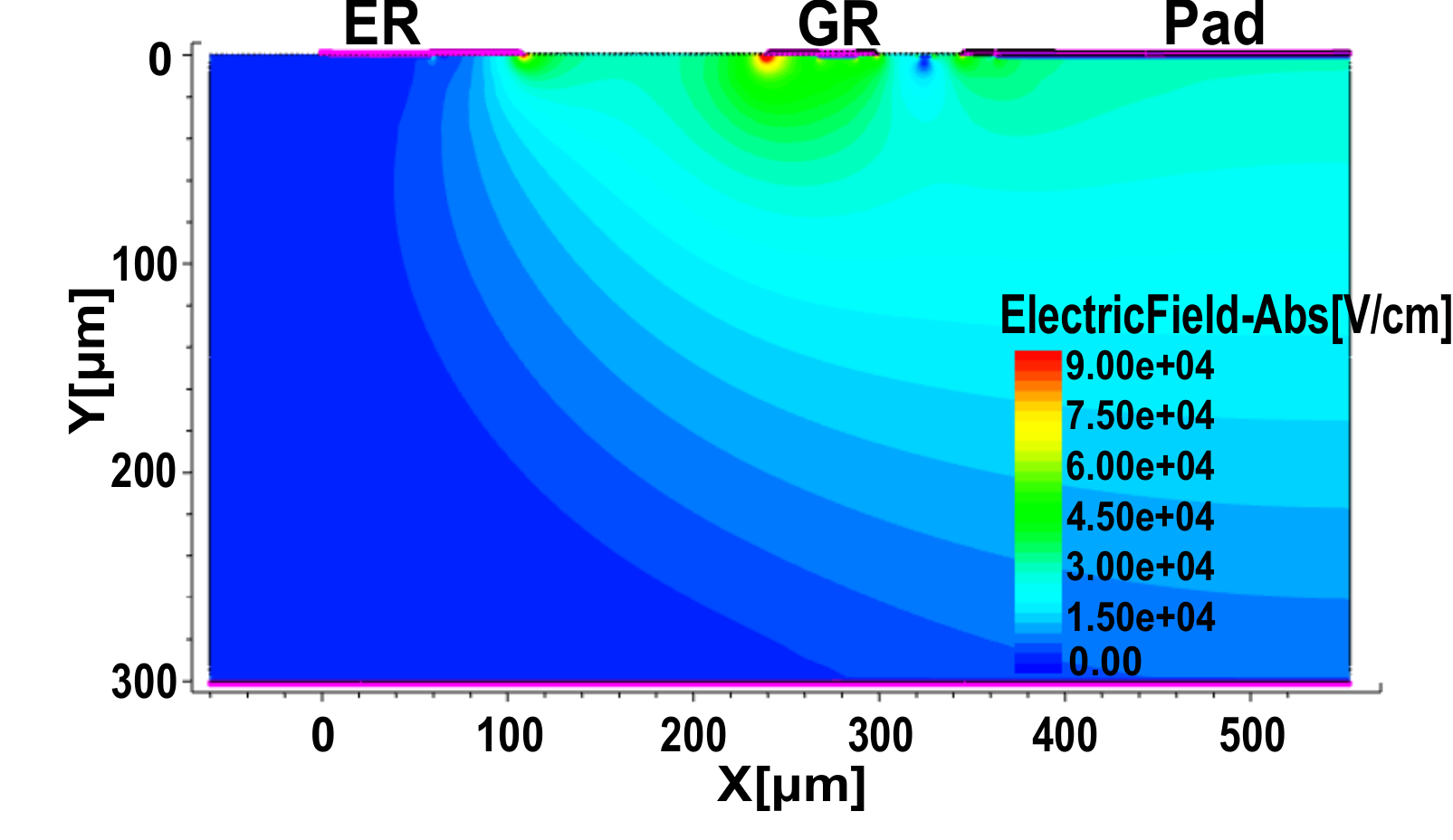}
}
\subfloat[Absolute electric field zoomed to the surface region]{\label{fig:efield_surface}
\centering
\hspace{0.42cm}
\includegraphics[height=.16\textheight, width=.3\textheight]{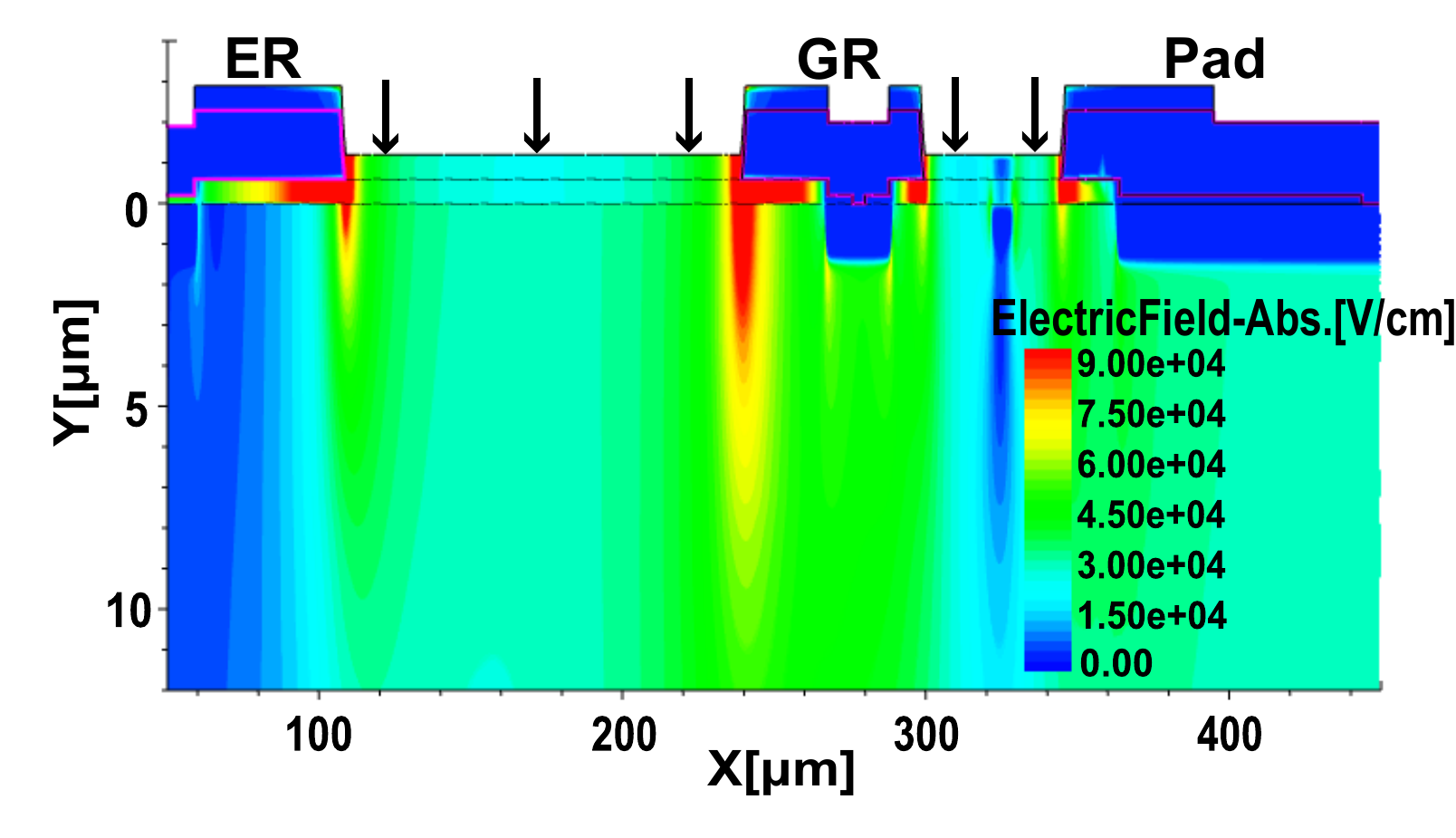}
}

\subfloat[Surface electric field parallel to the interface]{\label{fig:efield_x}
\centering
\includegraphics[height=.16\textheight, width=.3\textheight]{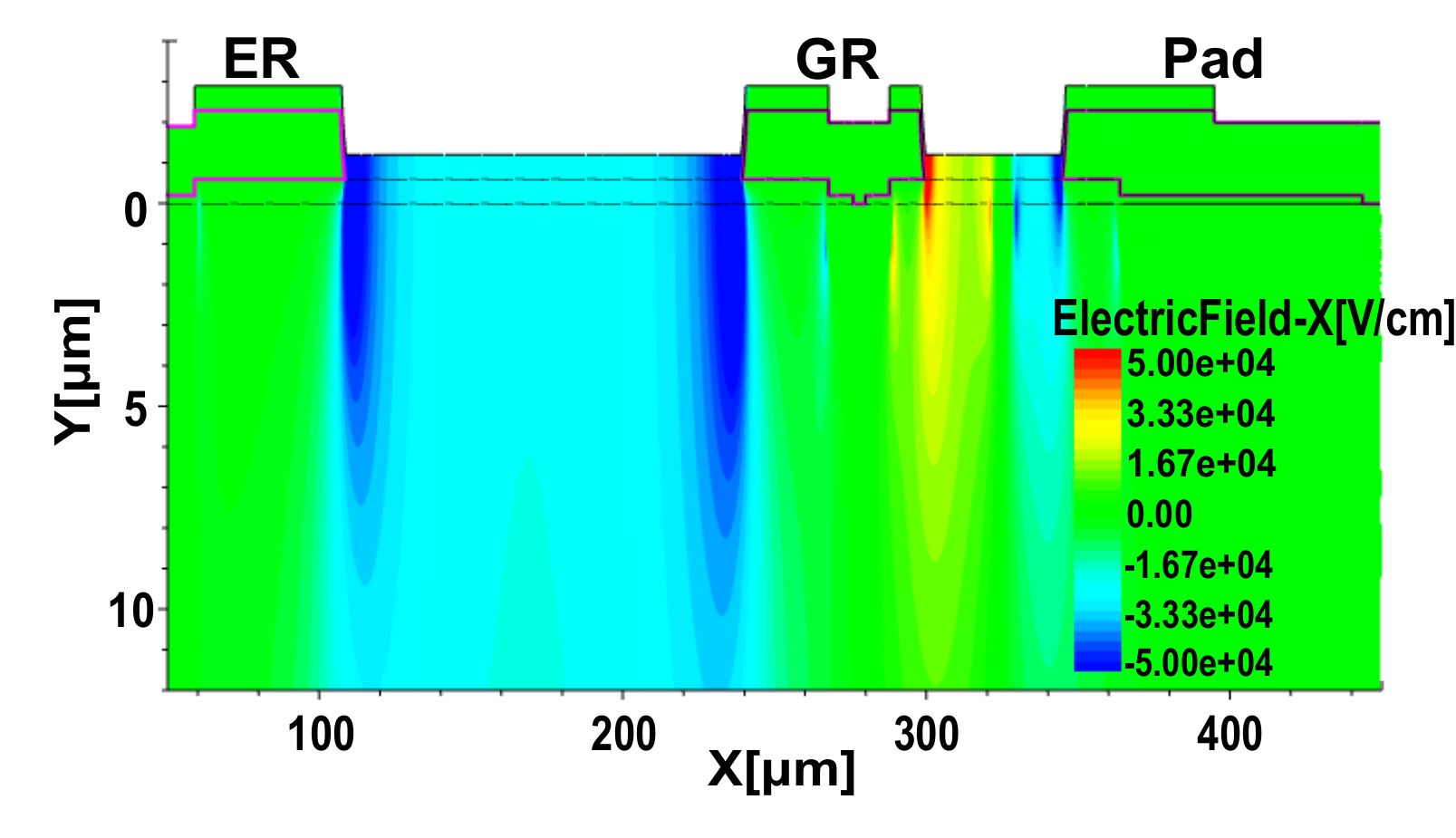}
}
\subfloat[Surface electric field orthogonal to the interface]{\label{fig:efield_y}
\centering
\hspace{0.47cm}
\includegraphics[height=.16\textheight, width=.3\textheight]{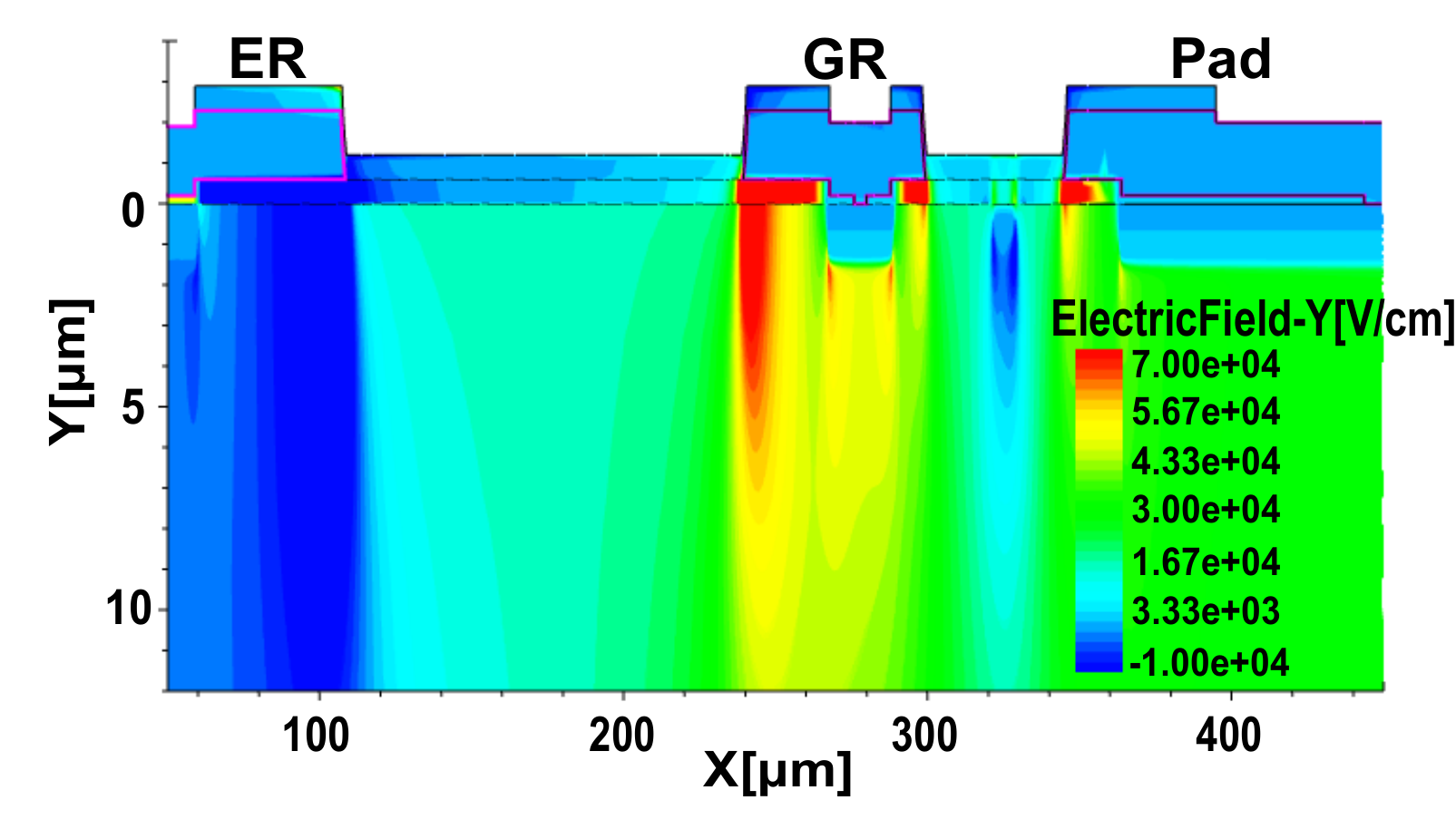}
}
\caption{\label{fig:tcad_efield} Maps of the electric field simulated with TCAD. The laser beam positions discussed in this paper are marked by arrows in~(b). The color scale is different for each plot.}
\end{figure}

The maps of the simulated electric field in the edge region at -500\,V bias are shown in Fig.~\ref{fig:tcad_efield}. Near the metal edges, the absolute electric field reaches values above 90\,kV/cm. The electric field parallel to the Si-SiO$_2$ interface between the ER and GR shows peaks at the metal edges and is relatively constant in between, as shown in Fig.~\ref{fig:efield_x}. In Fig.~\ref{fig:efield_y} a field pushing the electrons towards the interface can be observed in areas without metal covering, which is due to the positive fixed interface charge.

\subsection{Allpix$^2$ charge transport simulation}

Allpix$^2$~\cite{allpix} is a Monte Carlo simulation framework developed for semiconductor detectors. It can be combined with the electrostatic field, the static weighting field, and the doping profile simulated with TCAD, enabling faster charge transport simulations. 

To simulate charge transport at the sensor surface where a high electric field perpendicular to the surface exits, a new model called "surface reflectivity" was implemented in Allpix$^2$. It models the dynamics of charge carriers hitting the interface boundary as specular reflection with the component normal to the interface limited to 10~nm from the interface. This boundary condition accounts for charge carriers not being able to pass the potential barrier at the Si-SiO$_2$ interface. 
The parameters, geometry, and the mobility and multiplication models in Allpix$^2$ were the same as used in TCAD.

\begin{figure}[htbp]
\centering 

\subfloat[125\,$\upmu$m]{\label{fig:-150line}
\centering
\hspace{0.17cm}
\includegraphics[height=.121\textheight, width=.209\textheight]{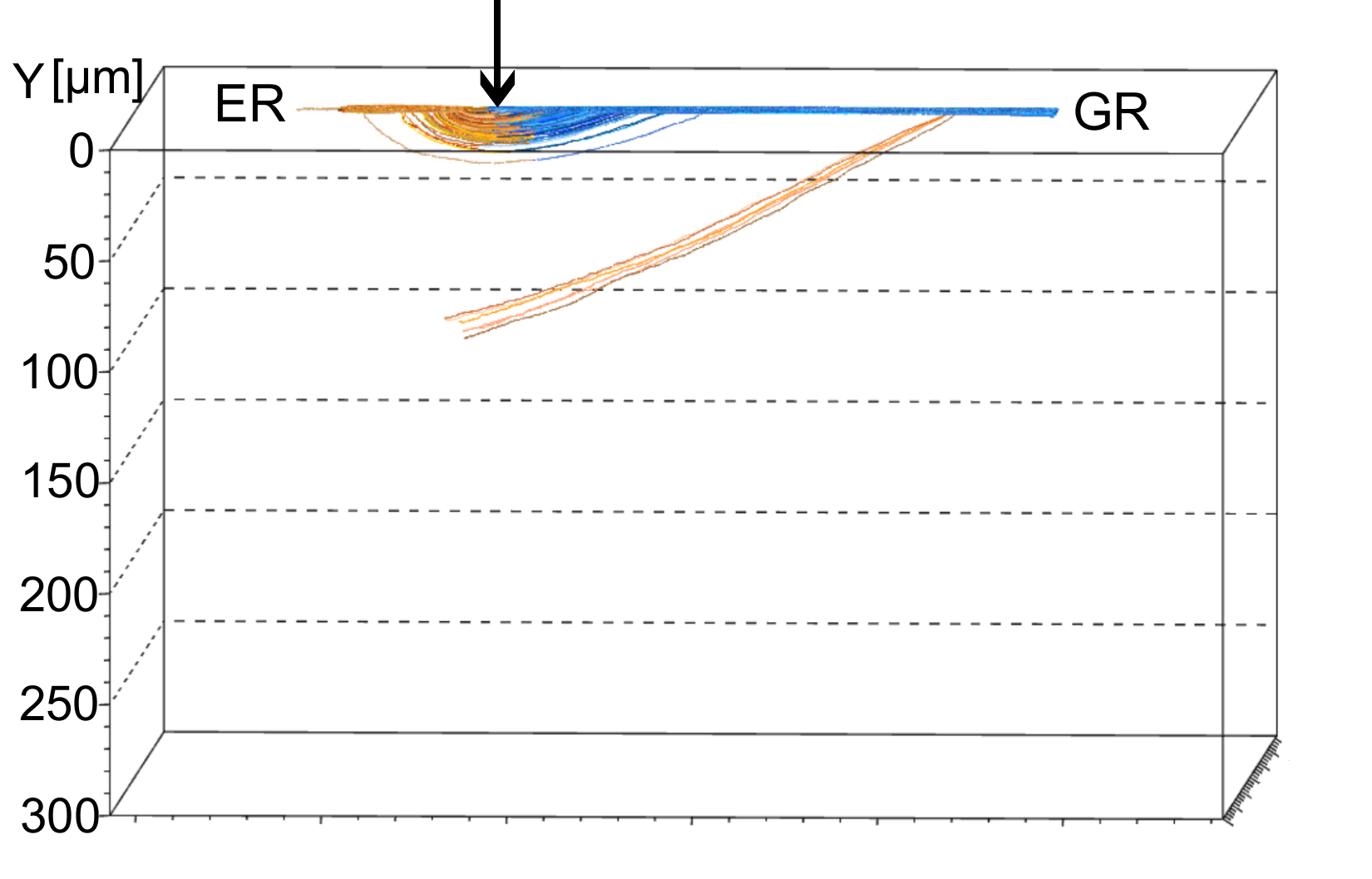}
}
\subfloat[175\,$\upmu$m]{\label{fig:-100line}
\centering
\hspace{0.12cm}
\includegraphics[height=.12\textheight, width=.21\textheight]{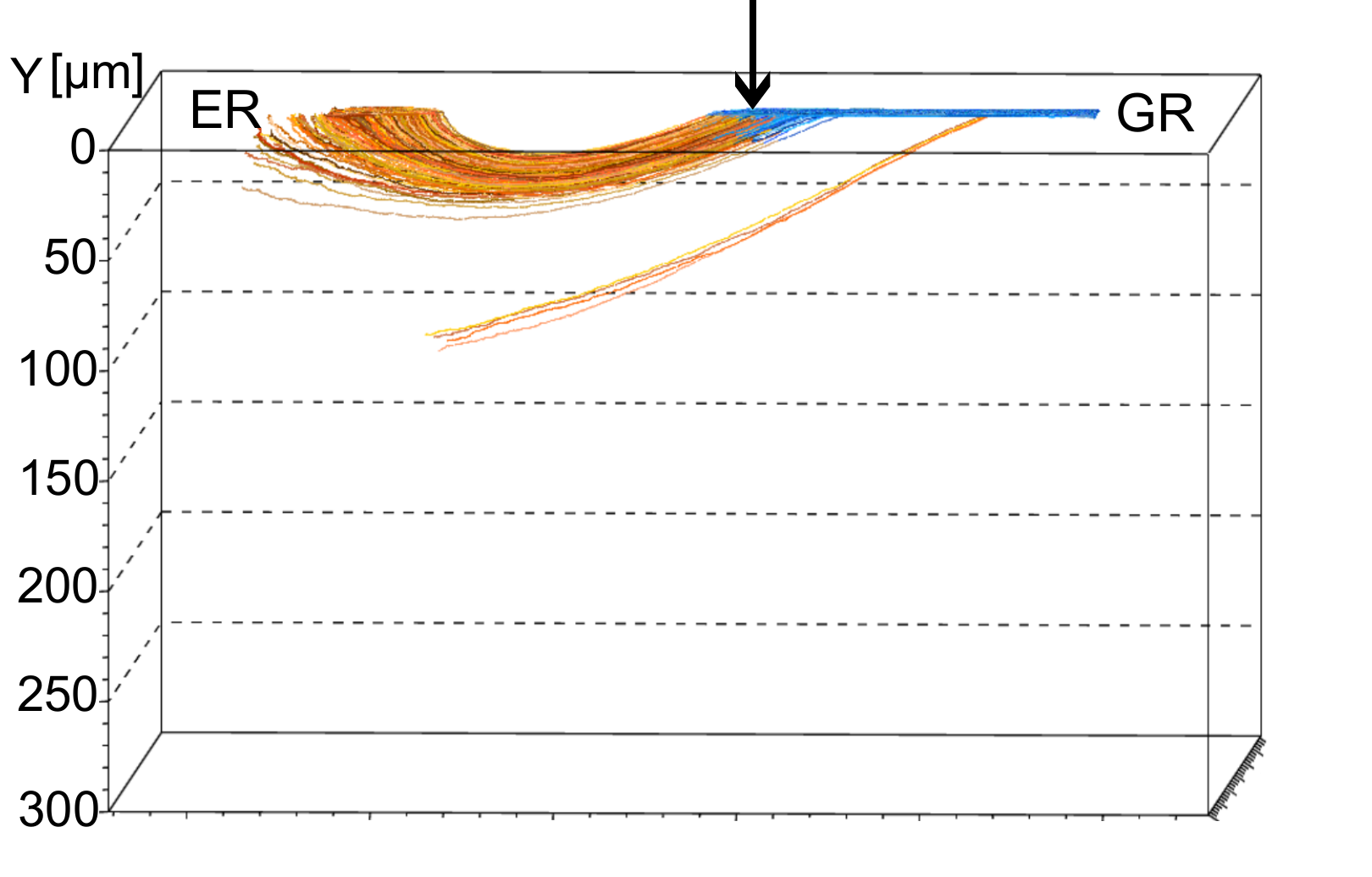}
}
\subfloat[225\,$\upmu$m]{\label{fig:-50line}
\centering
\hspace{0.1cm}
\includegraphics[height=.12\textheight, width=.212\textheight]{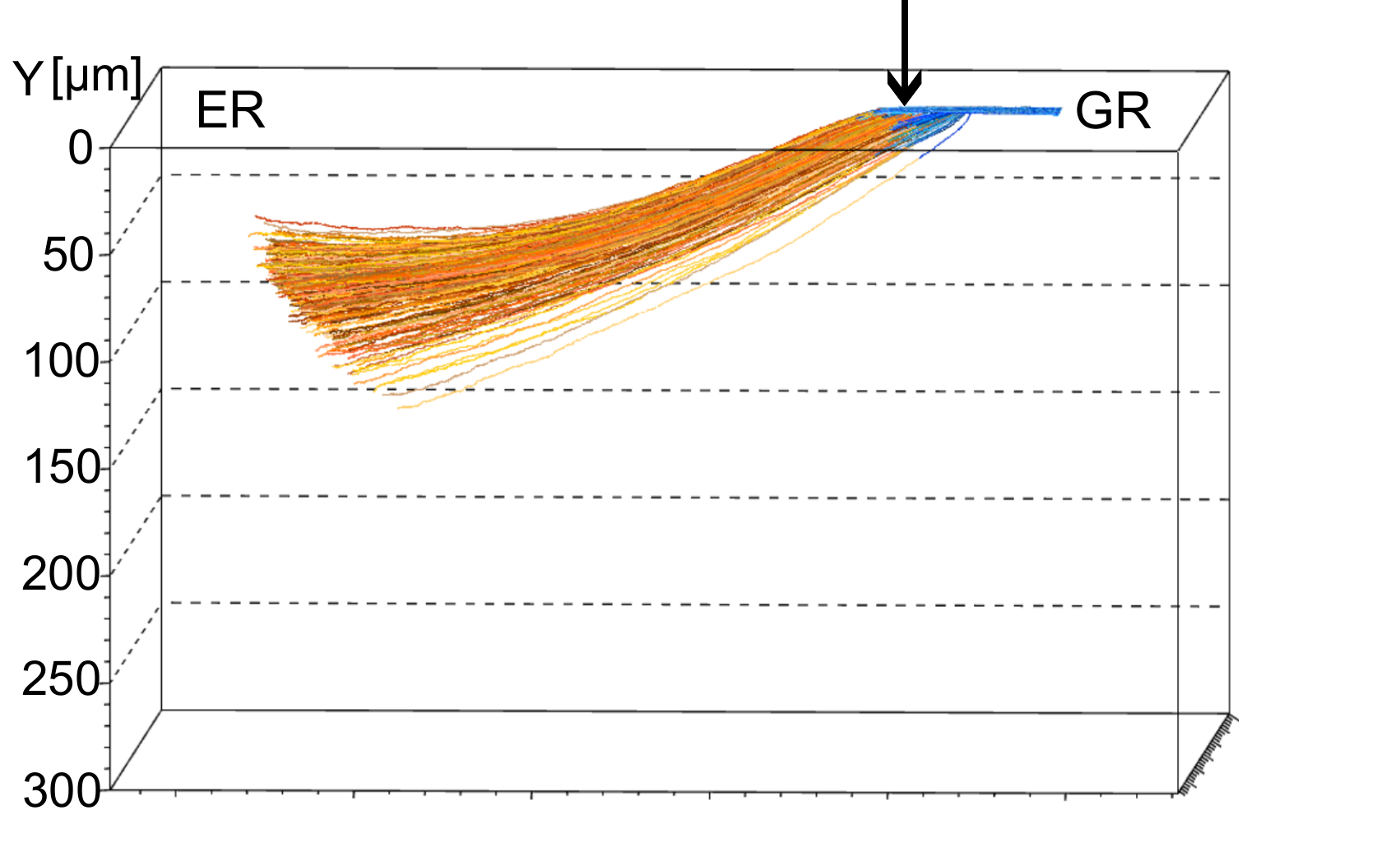}
}

\subfloat[125\,$\upmu$m]{\label{fig:155eh}
\centering
\includegraphics[height=.12\textheight, width=.215\textheight]{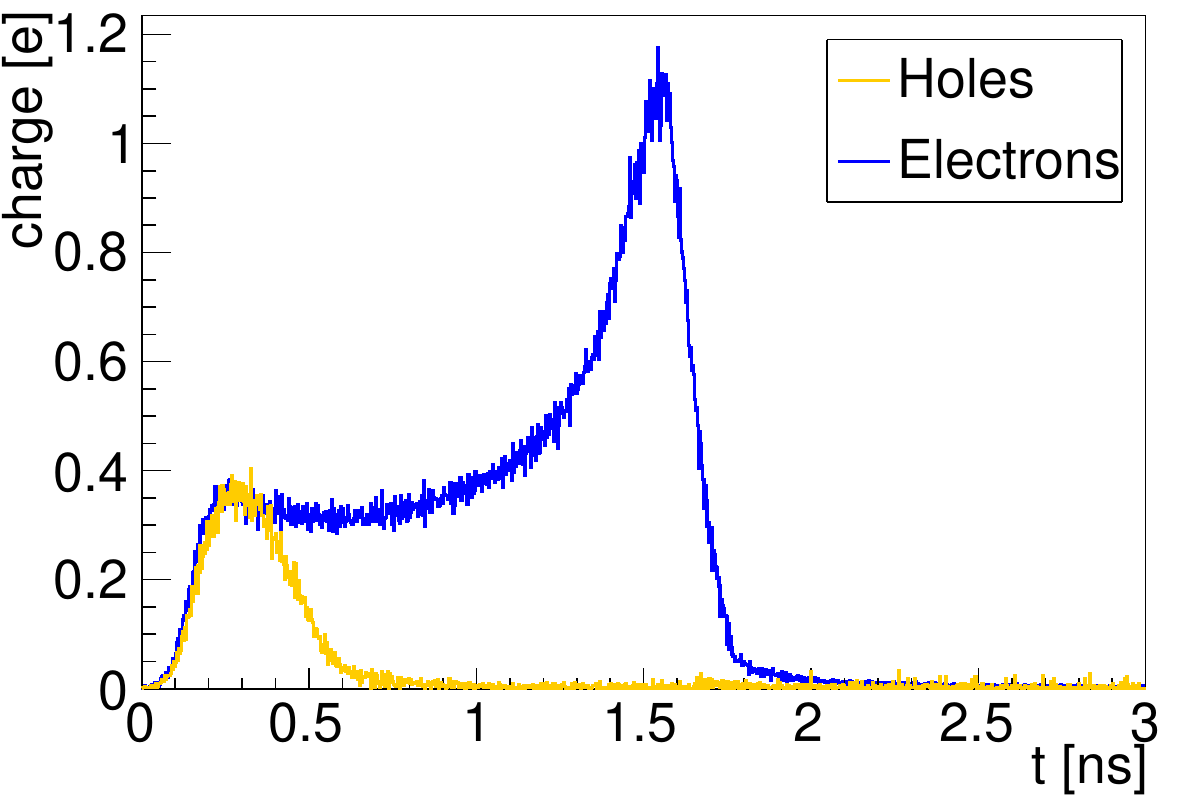}
}
\subfloat[175\,$\upmu$m]{\label{fig:105eh}
\centering
\hspace{0.22cm}
\includegraphics[height=.12\textheight, width=.2\textheight]{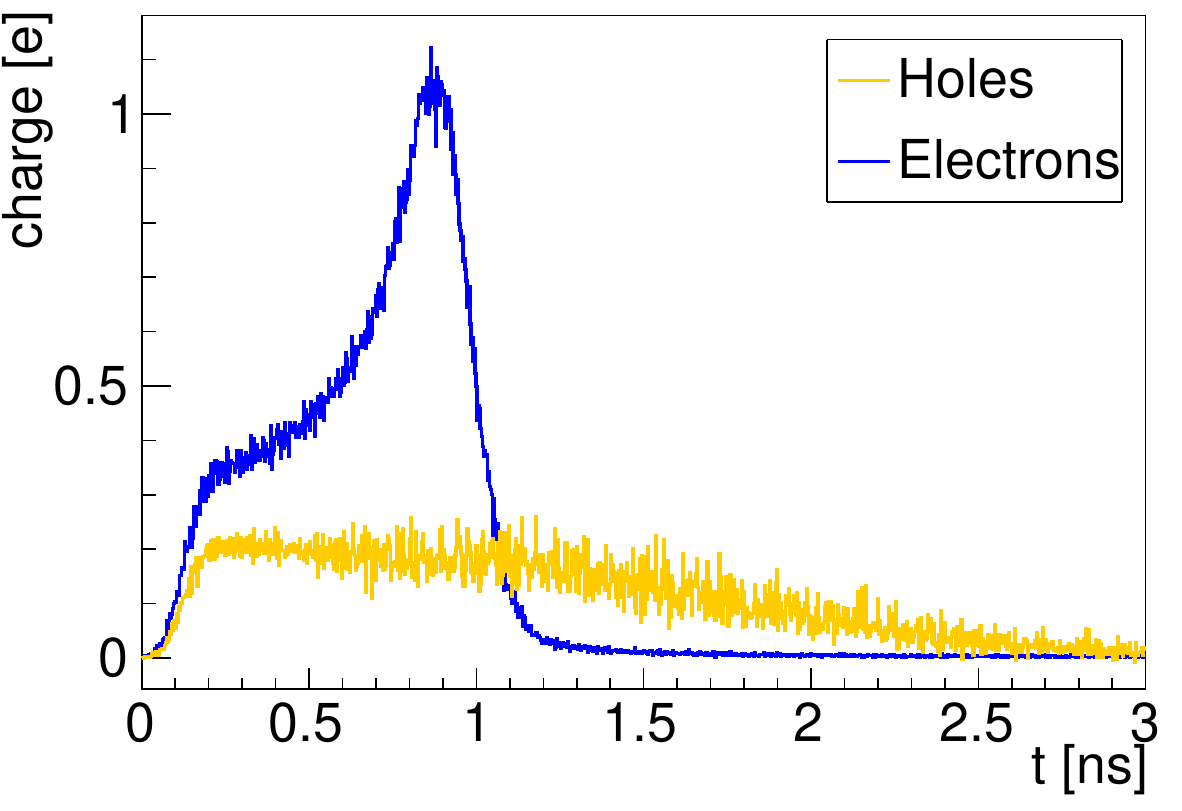}
}
\subfloat[225\,$\upmu$m]{\label{fig:55eh}
\centering
\hspace{0.22cm}
\includegraphics[height=.12\textheight, width=.2\textheight]{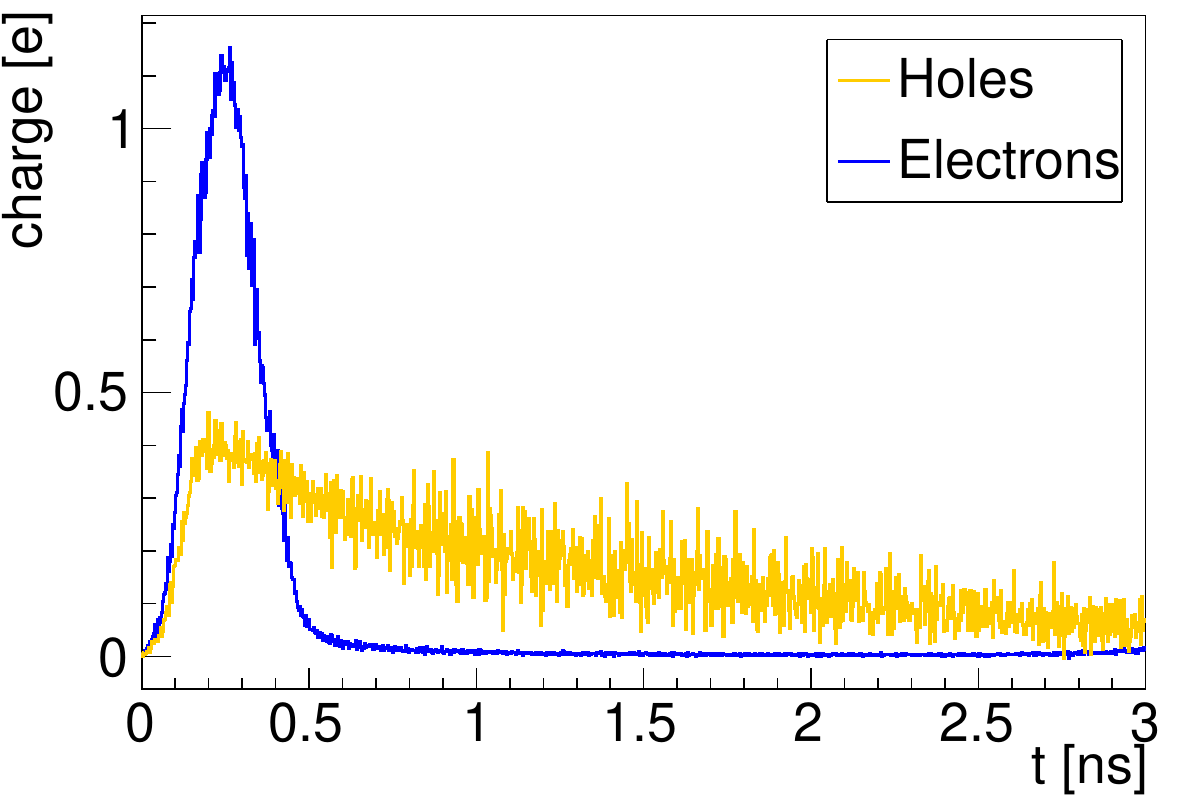}
}

\caption{\label{fig:linegraph}Contributions of electron and hole drift to the transient current from Allpix$^2$ simulation. The top row shows the paths of laser-injected carriers, the laser positions are indicated by arrows. The bottom row shows the induced charge per time step on the backside-ER electrode. }
\end{figure}

The simulated drift and diffusion of charge carriers and the resulting transients at 500\,V reverse bias for the selected laser positions are shown in Fig.~\ref{fig:linegraph}.
With the laser positioned near the ER ($x=125$\,$\upmu$m) holes are immediately accelerated towards the ER in the high-field region at the ER metal edge, resulting in the short hole signal in Fig.~\ref{fig:155eh}. Electrons, in contrast, drift a longer distance toward the GR. Once the electrons reach the high-field region at the GR metal edge, they induce a second peak due to faster drift velocity.
As the laser moves to the midpoint between the GR and the ER ($x=175$\,$\upmu$m) the hole peak disappears because the holes repulsed by the positive interface charge and avoid the electric field peak at the surface.
Near the guard ring ($x=225$\,$\upmu$m) the generated electrons immediately enter the high field of GR metal edge and get accelerated towards the electrode, inducing a sharp and short signal. All the holes drift towards the lateral edge of the diode.

\section{Comparison of Monte Carlo simulations, finite element simulations, and measurements}

In both measurements and simulations, a pulsed laser with a wavelength of 660\,nm was used to finely probe the edge region in 5~$\upmu$m increments along the $x$-axis. 
The measurements were conducted at approximately 50\% relative humidity. Since the laser scan was performed immediately after ramping the reverse bias voltage to 500\,V, it was assumed that surface ion migration was minimal~\cite{ninca2024tcad}, and thus the effect of humidity can be neglected.
The simulated induced transients were convoluted with the 
transfer function of the setup, which was determined by injecting known signals using a pulse generator~\cite{transfer}. 

\subsection{Transient currents and prompt current}

\begin{figure}[htbp]
\centering 
\hspace{-0.845cm}
\subfloat[125\,$\upmu$m]{\label{fig:-150}
\centering
\includegraphics[height=.155\textheight, width=.142\textheight]{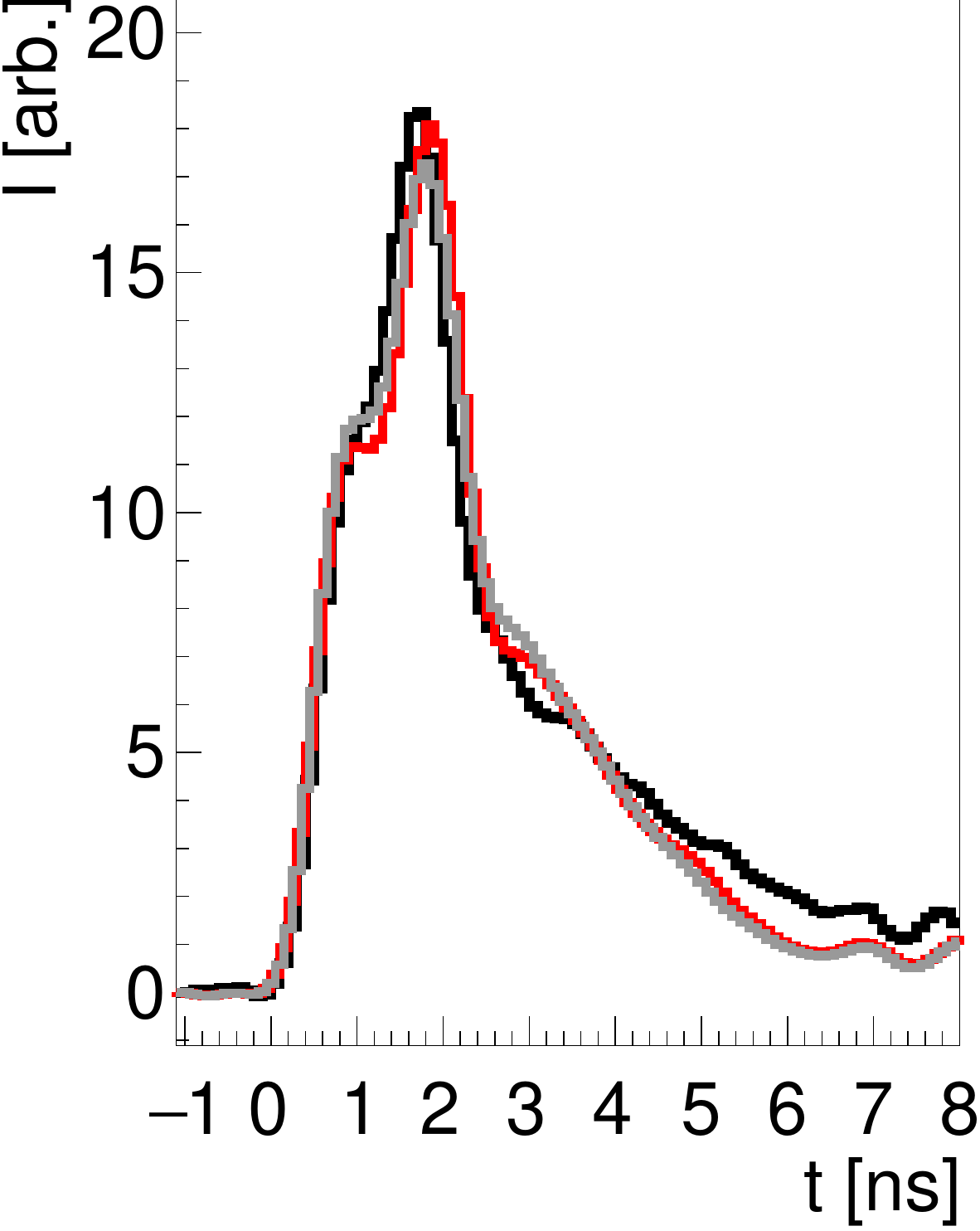}
}
\subfloat[175\,$\upmu$m]{\label{fig:-100}
\centering
\includegraphics[height=.155\textheight, width=.127\textheight]{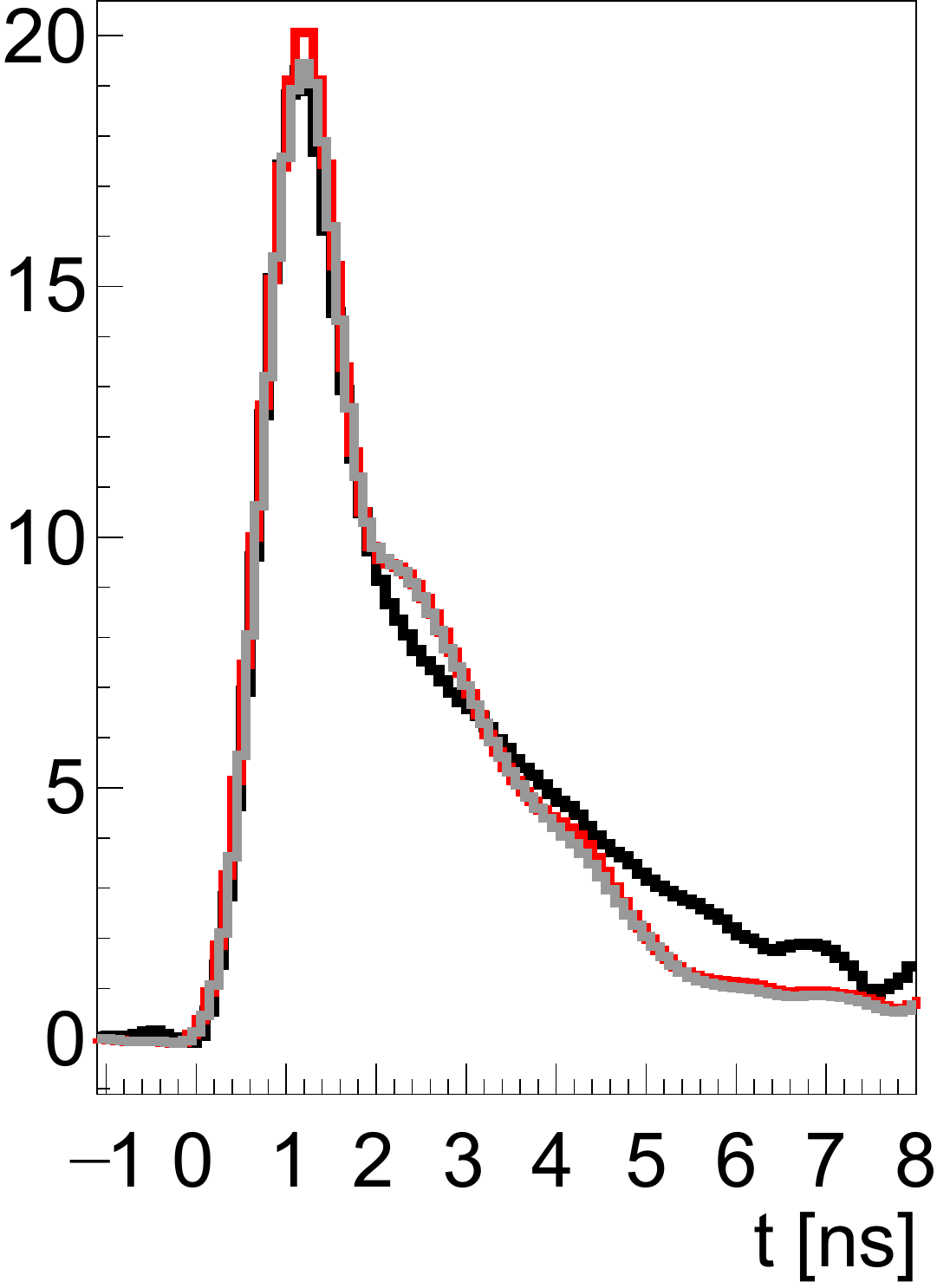}
}
\subfloat[225\,$\upmu$m]{\label{fig:-50}
\centering
\includegraphics[height=.155\textheight, width=.127\textheight]{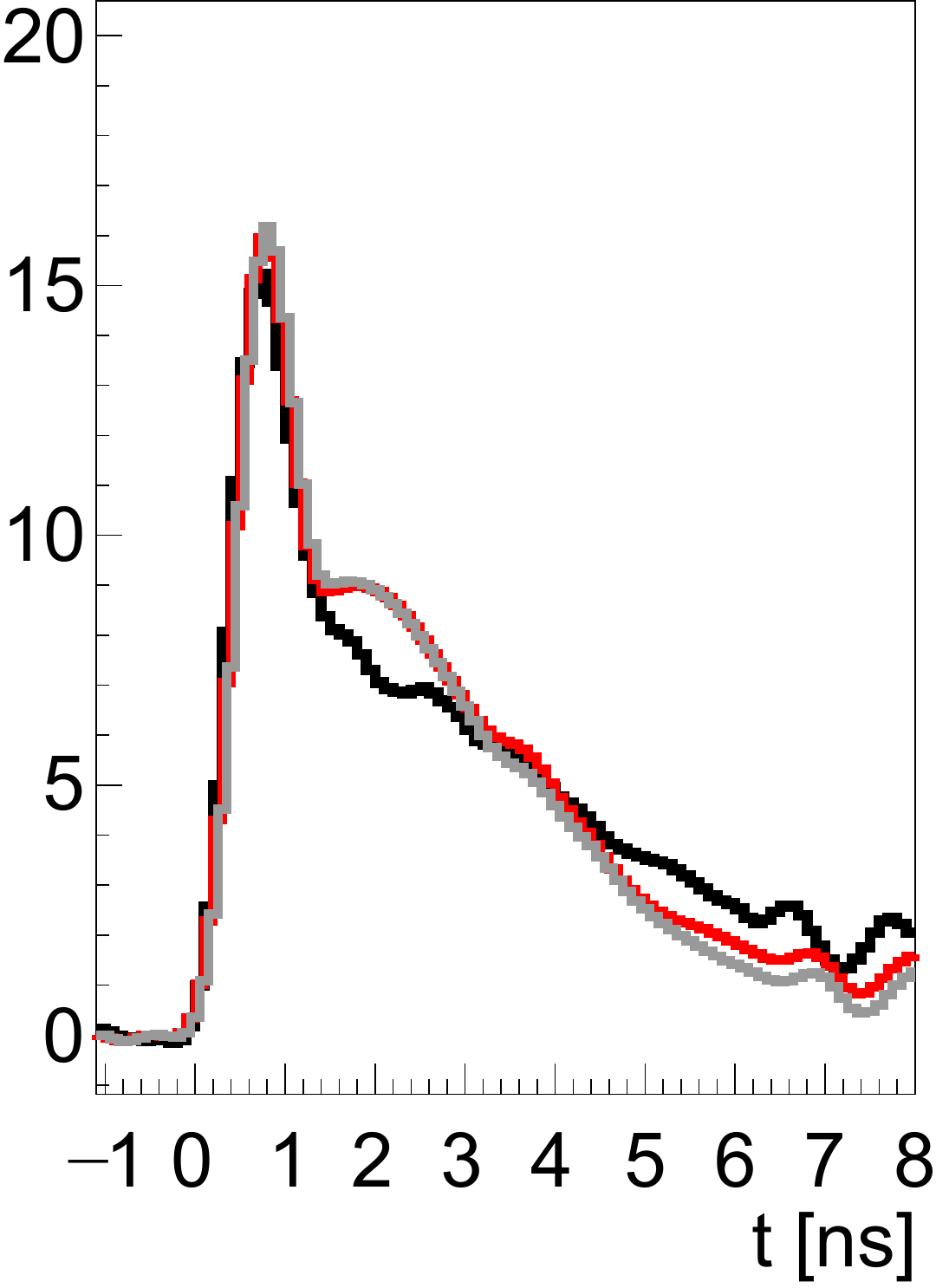}
}
\subfloat[305\,$\upmu$m]{\label{fig:30}
\centering
\includegraphics[height=.155\textheight, width=.127\textheight]{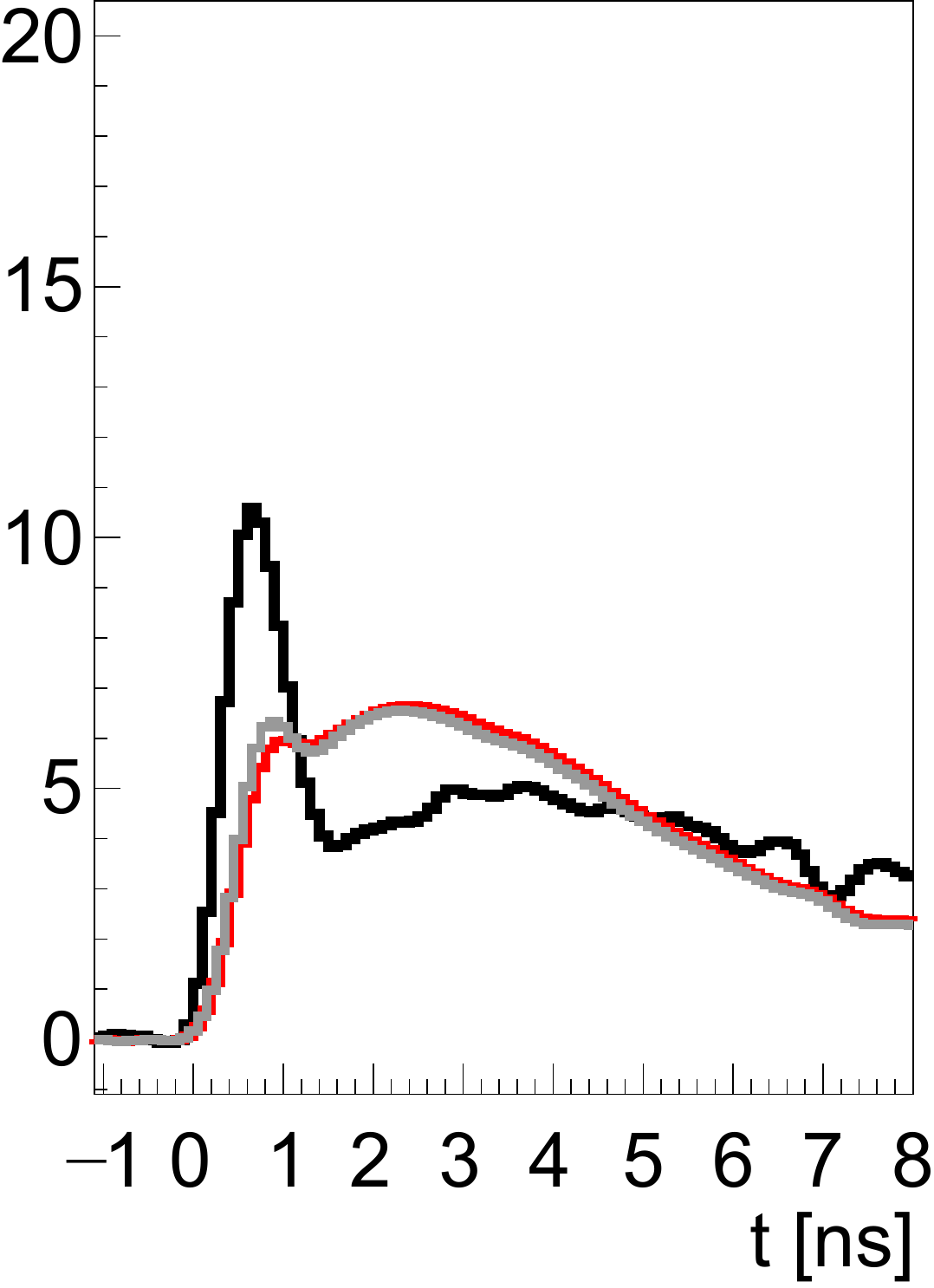}
}
\subfloat[335\,$\upmu$m]{\label{fig:60}
\centering
\includegraphics[height=.155\textheight, width=.142\textheight]{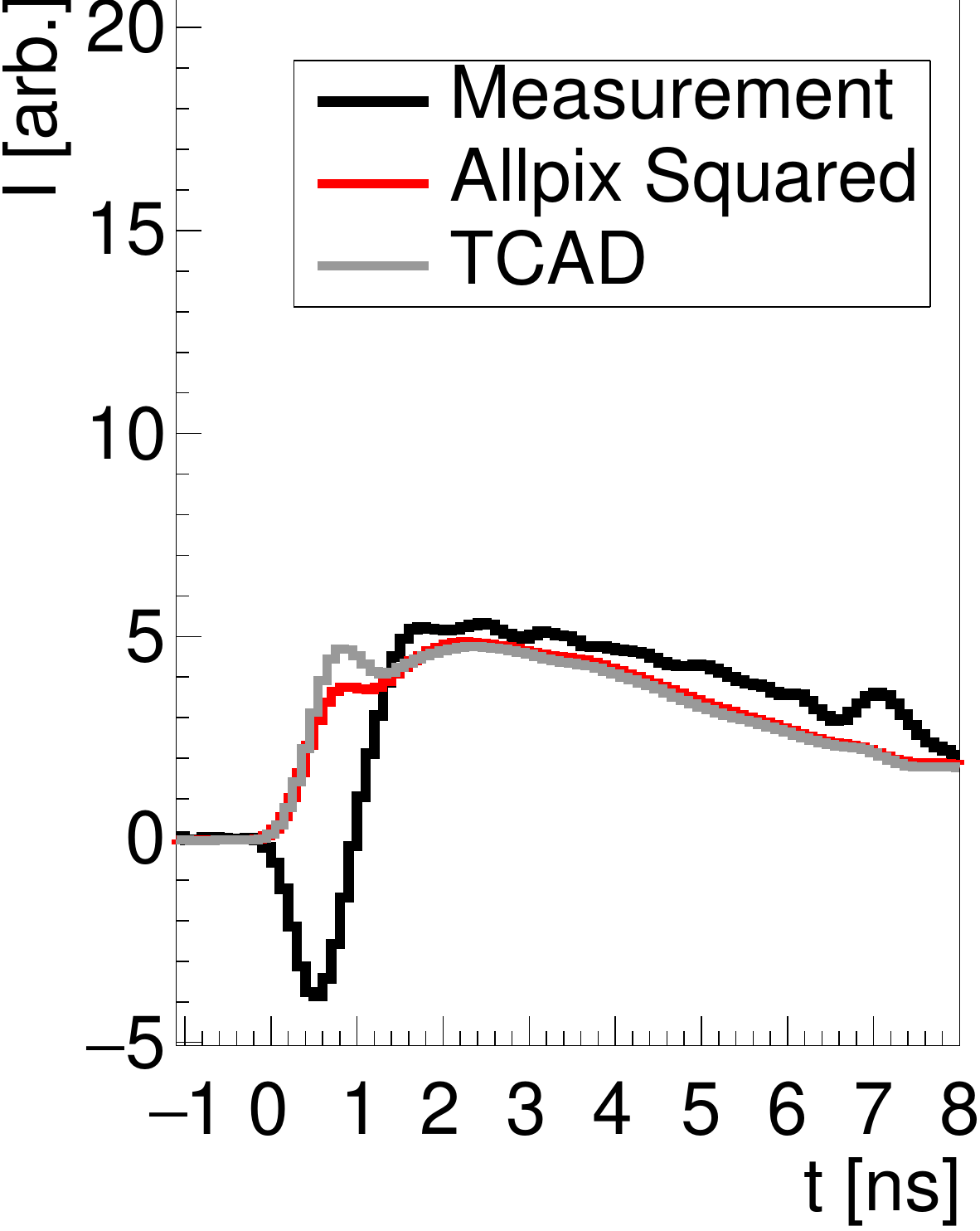}
}

\caption{\label{fig:transients2} Comparison of the measured transient currents (black) with Allpix$^2$ simulations (red) and TCAD simulations (grey) for selected positions across the full edge range. The simulations were scaled to match the measurements for the total induced charge in the region between the ER and GR (see Fig.~\ref{fig:charge_collection}).}
\end{figure}

The comparisons of the transient currents for selected laser positions are shown in Fig.~\ref{fig:transients2}. The simulations with Allpix$^2$ and with TCAD generally agree very well. Between the ER and the GR, the measured transients are well reproduced by the simulations (Fig.~\ref{fig:-150}-\ref{fig:-50}). The measured initial transient currents differ significantly from the simulations (Fig.~\ref{fig:30}-\ref{fig:60}) in the region between the GR and the pad.

To obtain information about the surface electric field as a function of the position of the diode, prompt current profiles were obtained by integrating over a short time interval ($\Delta t$ = 0.2\,ns, from $t=0.3$\,ns to $t=0.5$\,ns) of the rising edge of the transients. Based on previous measurements with the same diode, charge multiplication is not significant at a reverse bias of 500\,V. Therefore, the resulting prompt current $\sum_{\Delta t} I$ is proportional to $\mu(E) \cdot \vec{E}(x) \cdot \vec{E}_w(x)$.

The comparison of the prompt current profiles, calculated from the measured and simulated signals is shown in Fig.~\ref{fig:prompt_current}.
In both simulation and measurement, the maximum prompt current is observed near the GR. It steeply decreases as the laser beam moves away from the GR. The simulated prompt current near the GR is less than the measured, corresponding to the small discrepancy observed in the rising edge of the transients in Fig.~\ref{fig:-50}. This discrepancy suggests an underestimation of electric field in the simulation near the GR. It should be noted that the results could be influenced by an approximate mobility model used in simulations.
The large discrepancy of the negative signal observed in measurements between the GR and the pad electrode remains poorly understood. 

\subsection{Charge collection}

\begin{figure}[htbp]
\centering 
\subfloat[Prompt current profiles]{\label{fig:prompt_current}
\centering
\includegraphics[height=.175\textheight,width=.44\textwidth]{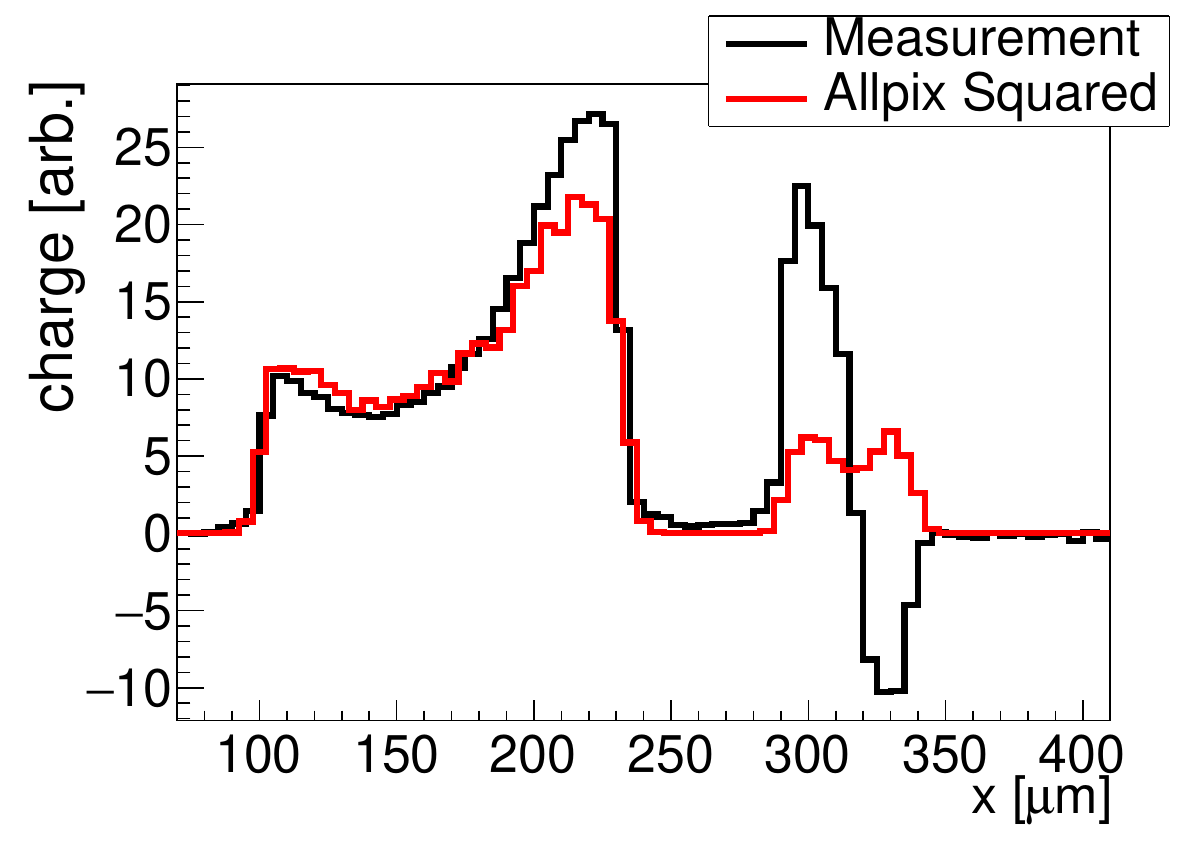}
}
\subfloat[Charge profiles]{\label{fig:charge_collection}
\centering
\includegraphics[height=.175\textheight, width=.44\textwidth]{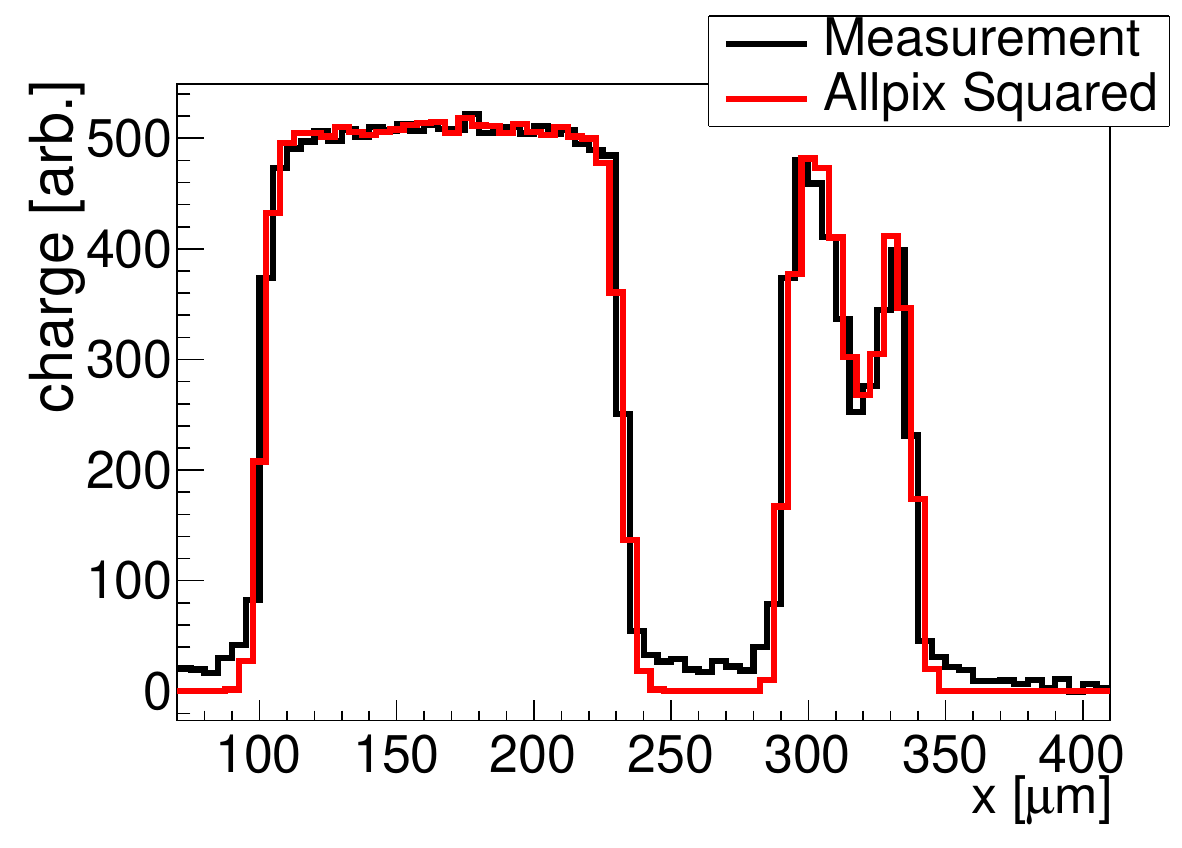}
}

\caption{Comparison of measured and simulated (a) prompt current and (b) charge profiles. There are no signals for $x<100$, $240<x<280$, $x>340$\,$\upmu$m because of the ER, GR, and pad metalizations, respectively. The simulations were scaled to match the measurements with the total induced charge in the region between the ER and GR.}
\end{figure}

The charge profile shows the integrals of the full transient current pulses as a function of the laser position.
As shown in Fig.~\ref{fig:charge_collection}, a relatively good agreement between simulation and measurement was achieved. The total collected charge remains uniform between the ER and GR, indicating no charge loss at the Si-SiO$_2$ interface. 
The reduction in collected charge observed between the GR and pad electrode can be explained by the presence of a potential well introduced by the p-stop (see Fig.~\ref{fig:efield_surface}). This also modifies the weighting field, resulting in a reduced hole signal. 
In diodes without p-stop, the charge collection remains constant between the GR and pad electrode, and it has the same amplitude as in the region between the ER and GR.

\section{Summary}

A new method for exploring high-gradient electric field region in the sensor periphery is proposed. It involves laser-based charge injection near the top surface and high-bandwidth recording of the resultant transient signals. The method was investigated by comparing test data with charge transport simulations.
A new parameter ”surface reflectivity” was implemented in Allpix$^2$ to model charge carrier transport in inversion or accumulation layers at the Si-SiO$_2$ interface as constrained specular reflections. Illumination of the periphery of a silicon sensor with 660\,nm focused laser pulses was simulated in TCAD and Allpix$^2$. A laser scan simulation over the edge region was performed in Allpix$^2$. The acquired data were analyzed as a function of the laser beam position in aspects of the transient pulses, prompt current and charge collection. Surface conditions and the surface electric field distribution were evaluated based on the comparison of simulation and measurement.

The comparison between simulation and measurement generally showed a good agreement for the region between the edge ring~(ER) and guard ring~(GR), showing the feasibility of combining the finite element simulation with a Monte Carlo simulation for validation of the simulation. Discrepancies in the propagation of injected charge carriers shortly after their generation were observed near a p-stop implant, suggesting problems in the modeling of the latter.

The results show that the proposed method can be used to study the electric field near the surface in the periphery of silicon sensors, which is important for validation and tuning of the parameters of a TCAD simulation. Additionally, the method can be used to study the evolution of the electric field with the bias voltage or over time at constant voltage, e. g. in humid conditions or to test different guard ring designs.

\acknowledgments

This research was supported by the German Federal Ministry of Education and Research (BMBF) as part of Verbundprojekt 05H2024. 

\vspace{1em}
\noindent {\large\textbf{Copyright}}

\vspace{0.8em}
2025 CERN for the benefit of the ATLAS ITk Collaboration. CC-BY-4.0 license.


\begin{thebibliography}{99}

\bibitem{okuto_crowell}
Y. Okuto and C. R. Crowell, \emph{Threshold energy effect on avalanche breakdown voltage in semiconductor junctions}, \emph{Solid-State Electronics}, {\bf 18.2} (1975), pg. 161-168

\bibitem{GR}
Y. C. Kao and E. D. Wolley, \emph{High-voltage planar p-n junctions}, \emph{Proceedings of the IEEE}, {\bf 55.8} (1967), pg. 1409-1414

\bibitem{ninca2024tcad}
I.-S. Ninca et al., \emph{TCAD simulations of humidity-induced breakdown of silicon sensors}, \emph{NIMA} {\bf 1067} (2024), pg. 169-729

\bibitem{123}
Y.~Unno et al., \emph{Specifications and pre-production of n$^{+}$-in-p large-format strip sensors fabricated in 6-inch silicon wafers, ATLAS18, for the Inner Tracker of the ATLAS Detector for High-Luminosity Large Hadron Collider}, \emph{JINST}, {\bf 18}, T03008 (2023) 




\bibitem{ramo}
S. Ramo, \emph{Currents induced by electron motion}, \emph{Proceedings of the IRE}, {\bf 27.9} (1939), pg. 584-585

\bibitem{wp}
W. Rieger,  \emph{An application of extensions of the Ramo–Shockley theorem to signals in silicon sensors}, \emph{NIMA}, {\bf 940} (2019), pg. 453-461

\bibitem{prompt_current}
G. Kramberger et al., \emph{Investigation of irradiated silicon detectors by edge-TCT}, \emph{IEEE Transactions on Nuclear Science}, {\bf 57} (2010), pg. 2294-2302

\bibitem{tcad}
Synopsys Inc., \emph{Synopsys Sentaurus TCAD}, (2025) \url{https://www.synopsys.com/manufacturing/tcad.html}

\bibitem{canali}
C. Canali et al., \emph{Electron and Hole Drift Velocity Measurements in Silicon and Their Empirical Relation to Electric Field and Temperature}, \emph{IEEE Transactions on Electron Devices}, {\bf 22 (11)} (1975), pg. 1045-1047

\bibitem{van}
R. Van Overstraeten, H. De Man, \emph{Measurement of the ionization rates in diffused silicon p-n junctions}, \emph{Solid-State Electronics}, {\bf 13.5} (1970), pg. 583-608

\bibitem{allpix}
S. Spannagel et al., \emph{Allpix$^2$: A modular simulation framework for silicon detectors}, \emph{NIMA}, {\bf 901} (2018), pg. 164-172


\bibitem{transfer}
C. Scharf and R. Klanner, \emph{Determination of the electronics transfer function for current transient measurements}, \emph{NIMA}, {\bf 779} (2015), pg. 1-5





\end{thebibliography}
\end{document}